\documentclass[twocolumn,prb]{revtex4}
\usepackage{amssymb}
\usepackage{graphicx}
\usepackage{dcolumn}
\usepackage{bm}
\begin{document}

\title[]{Equilibrium-charge diagram of single quantum dot in an axial magnetic field}

\author{Qing-Rui Dong }

\address{College of Physics and Electronics, Shandong Normal University, Jinan, Shandong, 250014, China}

\begin{abstract}
The chemical potential of a quantum dot parabolically confining one or two electrons is studied in an axial magnetic field. The number of
electrons in the dot is given by combination between the chemical potentials of the dots and leads. The equilibrium-charge diagram of the
dot is obtained in the phase space determined by the magnetic field and the lead voltage. The influence of the magnetic field on the
electron number and the electron transport is discussed. The dependence of the electron number on the magnetic fields and the lead voltage
is clear at a glance in the equilibrium-charge diagram. Moreover, the diagram is a compact tool to show the condition of the electron
transport.
\end{abstract}

\maketitle

\section{Introduction}
One quantum dot and two coupled quantum dots containing two conduction electrons define two-dimensional analogies to the $H$ atom and the
$H_{2}$ molecule. Such devices may serve as building blocks for future quantum processors\cite{Loss98,Li01,Bayer01,Koppens06}. Recent
advances in experimental techniques have made it possible to study quantum dots in the few-electron regime when quantum dots contain only
one or two conduction electrons\cite{Wiel03,Hatano05}. Based on the exact-diagonalization method, it was shown that the two-electron ground
state exhibits a phase transition from a singlet to a triplet state due to interactions\cite{Harju02,Helle05}. The electronic structure of
a laterally coupled two-electron quantum dot molecule is computed for different confinement strengths and for varying inter-dot separations
in external magnetic fields\cite{Popsueva07}. The stability diagram of a one- and two-electron quantum dot was calculated for larger
inter-dot separations in a related exponential double-well potential\cite{Zhang06}. In theses studies, the number of electrons in quantum
dots and whether tunnel resonance occurs are determined by combination between the chemical potentials of dots and leads, and the chemical
potentials of quantum dots can be adjusted by modulating gate voltages. Meanwhile, external magnetic fields are frequently applied to
control the electronic states and spins of quantum dots\cite{Harju02}, but whether the number of electrons in quantum dots is changed
unexpectedly by magnetic fields has been studied to a lesser extent. To analyze the dependence of chemical potentials on magnetic fields
allows one to estimate the effect of magnetic fields on the number of electrons in quantum dots.

In the paper, we numerically compute the chemical potential of single quantum dot parabolically confining one or two electrons in an axial
magnetic field. One- and two-electron Schr\"odinger equations are solved by exact diagonalization to obtain the chemical potentials as
functions of the magnetic field. The number of electrons in the dot is given by combination between the chemical potential of the dot and
leads. The equilibrium-charge diagram of the dot is obtained in the phase space determined by the magnetic field and the lead voltage. The
dependence of the electron number on the magnetic fields and the lead voltage is clear at a glance in the equilibrium-charge diagram.
Meanwhile, the diagram is a compact tool to show the condition of the electron transport. The result helps to control the electron number
of single quantum dot in external magnetic field and to use single quantum dot as a magnetic electron-transport switch.

\section{Model}

The number of electrons in a quantum dot is determined from the condition that the chemical potential of the dot is less than that of the
leads (left and right leads)\cite{Wiel03}. The linear regime of conductance is considered, implying $\mu_{L}-\mu_{R}\approx 0$, where
$\mu_{L}$ and $\mu_{R}$ are the chemical potentials of the left and right leads. The chemical potential of a quantum dot $\mu(N)$ is
defined as:\cite{Wiel03}
\begin{equation}\label{eqn:mu}
\mu(N)=E_G(N)-E_G(N-1),
\end{equation}
where $E_G(N)$ is the ground state energy of the $N$ electron state. One- and two-electron Schr\"odinger equations are solved by exact
diagonalization to obtain the ground state energy. The Hamiltonian describing two electrons parabolically confined in a two-dimensional
quantum dot is given by
\begin{equation}\label{eqn:two}
H(\textbf{r}_{1},\textbf{r}_{2})=h(\textbf{r}_{1})+h(\textbf{r}_{2})+C(\textbf{r}_{1},\textbf{r}_{2})+H_{Z},
\end{equation}
\begin{equation}\label{E1}
C(\textbf{r}_{1},\textbf{r}_{2})=\frac{e^2}{4\pi{\epsilon_{r}}{\epsilon_{0}}|\textbf{r}_{1}-\textbf{r}_{2}|},
\end{equation}
\begin{equation}\label{E1}
H_{Z}=g\mu_{B}\sum_{i}\textbf{B}{\cdot}\textbf{S}_{i},
\end{equation}
with the single-electron Hamiltonian $h(\textbf{r}_{i})$ given as
\begin{equation}\label{eqn:single}
h(\textbf{r})=\frac{1}{2m^{*}}(\textbf{p}+e\textbf{A})^{2}+\frac{1}{2}m^{*}\omega_{0}^{2}\textbf{r}^{2},
\end{equation}
\noindent Here, $m^*=0.067m_e$ is the electron effective mass, $\epsilon_{r}=12.4$ is the dielectric constant, $g=-0.44$ is the g-factor in
GaAs, $\mu_B$ is the Bohr magneton, ${\bf A} = \frac{1}{2}[-By,Bx,0]$ is the vector potential for the magnetic field $B$ oriented
perpendicular to the $xy$ plane and the confinement strength $\hbar\omega_{0}$ = 3.0 meV. The calculation is performed at the finite
magnetic field strength (0$\backsim$6 T), which is the typical strength used in qubit operations.

The single-electron Hamiltonian Eq. (\ref{eqn:single}) and the two-electron Hamiltonian Eq. (\ref{eqn:two}) are diagonalized numerically to
obtain the ground state eigenvalues which are used to compute the chemical potentials $\mu(1)$ and $\mu(2)$ according to Eq. (\ref{eqn:mu})
[note that $E_G(0)\equiv 0$]. Throughout the work, the lowest 8 two-dimensional harmonic states have been used as the set of basis wave
functions and the convergence has been checked.

\section{chemical potential of single quantum dot}
\begin{figure}[htp]
\includegraphics[height=7cm]{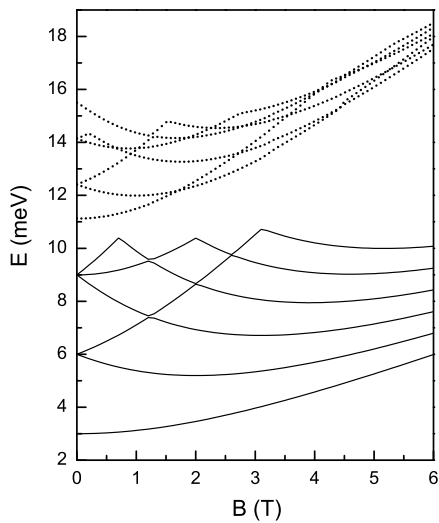}
\caption{\label{fig:flevels} The solid lines are six lowest single-electron energy levels as functions of the magnetic field, and the
dotted ones are six lowest two-electron energy levels as functions of the magnetic field. The single-electron levels in a parabolic dot are
well-known Fock-Darwin levels\cite{Darwin30}.}
\end{figure}
\begin{figure}[htp]
\includegraphics[width=6cm]{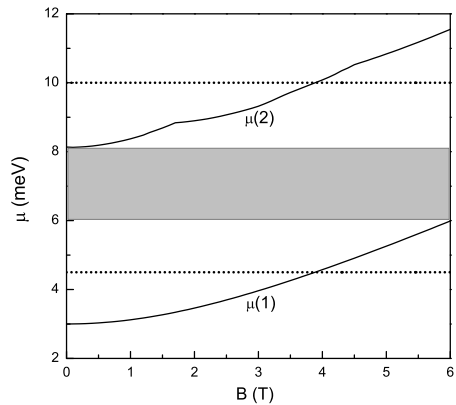}
\caption{\label{fig:fmu} The chemical potential of the dot $\mu(1)$ and $\mu(2)$ vs the magnetic fields $B$ (0$\sim$6 T). The gray zone
indicates a gap of the chemical potential, where the chemical potential of the quantum dot can not choose a value by adjusting magnetic
fields. The above horizontal line indicates the value of the chemical potential of the leads at which the schematic diagram in Fig.
\ref{fig:ChemialP}(a) is drawn. The below horizontal line is for Fig. \ref{fig:ChemialP}(b)}
\end{figure}
Fig. \ref{fig:flevels} displays the dependence of the energy levels $E(1)$ and $E(2)$ for one and two electrons on the magnetic field. For
the single-electron levels (well-known Fock-Darwin levels\cite{Darwin30}), at low values of the magnetic field, there is a hybridization of
Landau levels with the levels that arise from spatial confinement. As the magnetic field becomes stronger, the effect of the magnetic field
prevails over that of spatial confinement and thus the levels trend to the Landau-type levels. For two-electron states, the
electron-electron interaction energy gives a great contribution to the energy of two-electron states\cite{Harju02}. Here, we only care the
ground states, which determines the chemical potentials of the dot. In contrast to $E_G(1)$, the graph of $E_G(2)$ as a function of the
magnetic field is not smooth, which results from the effect of the interaction between two electron. Based on these analysis, it is easy to
explain that the dependence of the chemical potential of the dot on the magnetic field (Fig. \ref{fig:fmu}) according to Eq.
(\ref{eqn:mu}). The dependence of $\mu(1)$ on the magnetic field lies on $E_G(1)$ due to $E_G(0)\equiv 0$. The chemical potential $\mu(2)$
increases irregularly as the magnetic field becomes stronger, which results from the dependence of the difference between $E_G(2)$ and
$E_G(1)$ on the magnetic field.

\begin{figure}[htp]
\includegraphics[width=8cm]{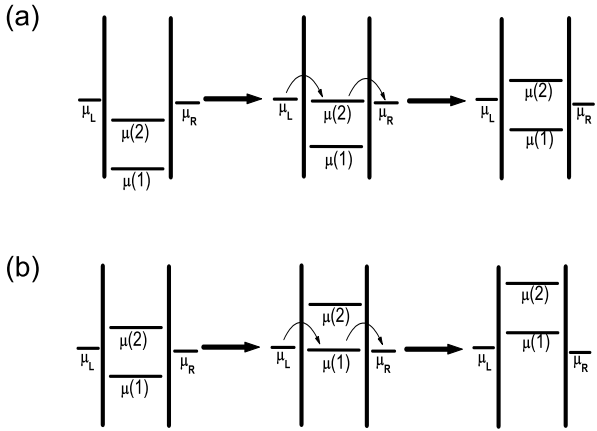}
\caption{\label{fig:ChemialP} Schematic diagrams of the chemical potentials in the dot and leads in the linear regime: (a) displaying the
process of reducing the electron number from $2$ to $1$ and the electron transport at the point $\mu(2)$ = $\mu_{L}$($\mu_{R}$). (b)
displaying the process of reducing the electron number from $1$ to $0$ and the electron transport at the point $\mu(1)$ =
$\mu_{L}$($\mu_{R}$). }
\end{figure}

In the following, the number of electrons in the dot is given by analyzing the variation of the chemical potential with the magnetic field.
When the chemical potential of the leads $\mu_{lead}$ is between 8 meV and 11.5 meV and are kept unchanged, $\mu(2)$ lows at $\mu_{lead}$
in absence of magnetic fields. In this case, two electrons are confined in the dot (see Fig. \ref{fig:ChemialP}(a)). As the magnetic field
becomes stronger, $\mu(2)$ is equal to $\mu_{lead}$ at a certain point and a resonant conductance is found (Fig. \ref{fig:ChemialP}(a)).
The electron transport must satisfy the condition that the chemical potential $\mu(2)$ is between $\mu_{L}$ and $\mu_{R}$. As the magnetic
field becomes strong further, electrons can not enter the dot from the left lead, and the electron transport is stopped (Fig.
\ref{fig:ChemialP}(a)). Moreover, one electron in the dot is lost by the right lead and the number of electrons in the dot is $1$.

When $\mu_{lead}$ is between 8 meV and 11.5 meV, the other electron in the dot can not be removed by increasing $B$ up to 6 T . The reason
is that there exists a gap in the chemical potential (see Fig. \ref{fig:fmu}). When $\mu_{lead}$ is between 3 meV and 6 meV, the number of
electrons in the dot can be changed from $1$ to $0$ as the magnetic field increases. In absence of magnetic fields, $\mu(1)$ lows at
$\mu_{lead}$ and one electron is confined in the dot (see Fig. \ref{fig:ChemialP}(b)). As the magnetic field becomes stronger, $\mu(1)$ is
equal to $\mu_{lead}$ at a certain point and a resonant conductance is found (Fig. \ref{fig:ChemialP}(b)). As the magnetic field becomes
strong further, the electron transport is closed and the number of electrons in the dot becomes $0$ (Fig. \ref{fig:ChemialP}(b)).


\section{Equilibrium-charge diagram of single quantum dot}

\begin{figure}[htp]
\includegraphics[width=7cm]{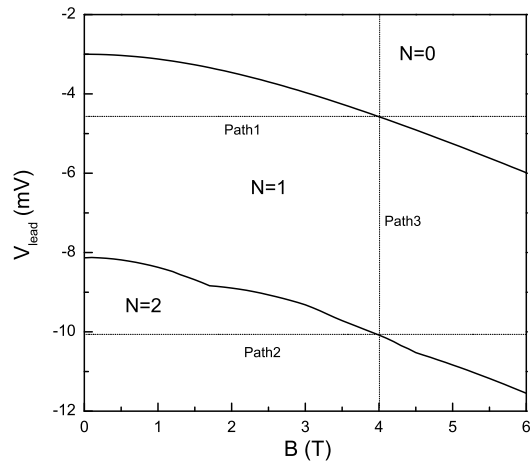}
\caption{\label{fig:stability} The equilibrium-charge diagram of the dot is plotted in ($B$, $V_{lead}$)-phase space with three domains of
the electron number $N = 0, 1, 2$. The solid lines indicate the boundaries between the domains; The dotted lines indicate three different
paths, which is used to show three typical evolution of the electron number.}
\end{figure}

Although the analysis of chemical potentials can explain physically the influence of magnetic fields on the electron number, the detailed
evolution of the electron number is not clear at a glance. Moreover, the electron number is determined by combination between the chemical
potentials of the dot and leads. Thus, the electron number can also be modulate by adjusting the chemical potential of the leads, which is
related to the lead voltage $V_{lead}$ and can be expressed by $\mu_{lead} = - |e|V_{lead}$. In the background, it is useful to plot a
equilibrium-charge diagram in ($B$, $V_{lead}$)-phase space. In Fig. \ref{fig:stability}, the equilibrium-charge diagram of the dot is
plotted in ($B$, $V_{lead}$)-phase space with three domains (the electron number $N = 0, 1, 2$). Once provided a certain value ($B$,
$V_{lead}$), one can obtain a certain electron number by the diagram. The diagram is rather similar to Fig. \ref{fig:fmu}, while their
suggestions are quite distinct from each other. In fact, The diagram is a transformation of Fig. \ref{fig:fmu}, which describes the
dependence of the chemical potential of the dot on the magnetic field. The transformation involves that the $xy$ plane represents the ($B$,
$V_{lead}$)-phase space instead of a functional relation. As a consequence of the transformation, the lines of the diagram indicate the the
boundaries between the different domains while the lines present the chemical potential of the dot in Fig. \ref{fig:fmu}. The change can be
explained by the fact that the electron number is determined by combination between the chemical potentials of the dot and leads.

\begin{figure}[htp]
\includegraphics[width=7cm]{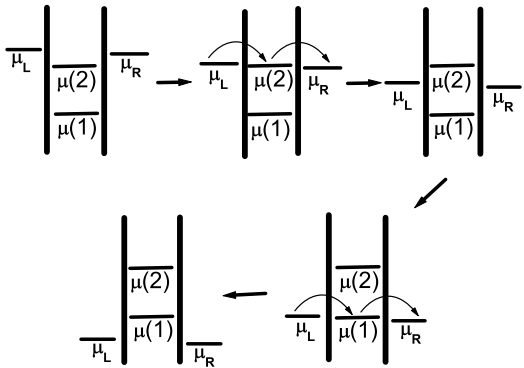}
\caption{\label{fig:schematicV} Schematic diagrams of the chemical potentials in the dot and leads in the linear regime: displaying the
process of reducing the number of electrons in the dot from $2$ to $1$ and then to $0$. The electron transport is turned on at the point
$\mu(2)$ = $\mu_{L}$($\mu_{R}$) and at the point $\mu(1)$ = $\mu_{L}$($\mu_{R}$). }
\end{figure}

In the following, we analyze the evolution of the electron number in ($B$, $V_{lead}$)-phase space by three typical examples. In these
examples, the electron number changes along three different paths (Fig. \ref{fig:stability}). As the magnetic field increases along $Path
1$, the system enter the domain $N=0$ from the domain $N=1$, which indicates the electron number is reduced from $1$ to $0$. In this
process, the voltage of the leads keeps unchanged, thus, the chemical potential of the leads also keeps unchanged and the chemical
potential of the dot increases, which corresponds to Fig. \ref{fig:ChemialP}(b). When $Path 1$ meets with the boundary between the domains,
the chemical potential of the dot is equal to that of the leads and a resonant conductance is found. Similarly, as the magnetic field
increases along $Path 2$, the system enter the domain $N=1$ from the domain $N=2$, which indicates the electron number is reduced from $2$
to 1. The chemical potential of the leads also keeps unchanged and the chemical potential of the dot increases, which corresponds to Fig.
\ref{fig:ChemialP}(a). When $Path 2$ meets with the boundary between the domains, the chemical potential of the dot is equal to that of the
leads and a resonant conductance is found. As the voltage of the leads increases along $Path 3$, the system enter the domain $N=1$ from the
domain $N=2$ and then enter the domain $N=0$ from the domain $N=1$, which indicates the electron number is reduced from $2$ to $1$ and then
from $1$ to $0$. In this process, the magnetic field keeps unchanged, thus, the chemical potential of the dot also keeps unchanged and the
chemical potential of the lead decreases. Schematic diagrams of the chemical potentials for this process is given in Fig.
\ref{fig:schematicV}. $Path 3$ meets two times with the boundaries and thus the electron transport occurs two times. The three typical path
can represent the revolution of the electron number in ($B$, $V_{lead}$)-phase space.

The analysis above show that the number of electrons in the dot can be modulated by adjusting combination between the magnetic fields and
the lead voltage. In general, the electron number decreases as $V_{lead}$ increases and decreases as $B$ becomes stronger. The
equilibrium-charge diagram provides the whole evolution of the electron number in ($B$, $V_{lead}$)-phase space. The result helps to guide
external magnetic field as a tool to modulate the number of electrons with the aid of the voltage of the leads. On the other hand, the
result helps to avoid that external magnetic field change unexpectedly the number of electrons when the magnetic field is used for other
aims. Moreover, the electron transport must satisfy the condition that the chemical potential of the dot is between $\mu_{L}$ and $\mu_{R}$
and all points on the boundaries between the domains satisfy this condition, so resonant conductances can be found on the boundaries
between the domains. Thus, the equilibrium-charge diagram is a compact tool to show clearly the condition of the electron transport. The
result helps to control the electron transport by adjusting external magnetic field and helps to use single quantum dot as a magnetic
electron-transport switch. The transport through the switch is turned on when the point ($B$, $V_{lead}$) move to the boundaries, and the
transport is turned off when the point ($B$, $V_{lead}$) is out of the boundaries.

\section{Conclusion}

Usually electron transport properties of quantum dots is studied by adjusting gate voltages, which corresponds to the $\omega_{0}$ in Eq.
(\ref{eqn:single}). In this paper, the chemical potentials of the dot are modulated by means of adjusting external magnetic fields, which
corresponds to the $A$ in Eq. (\ref{eqn:single}). The equilibrium-charge diagram of the dot is obtained in the phase space determined by
the magnetic field and the lead voltage. The dependence of the number of electrons on the magnetic fields and the lead voltage is clear at
a glance in the equilibrium-charge diagram. The result shows external magnetic field can be used as a tool to modulate the number of
electrons of single quantum dot with the aid of the voltage of the leads. On the other hand, the result suggests that the number of
electron can be changed unexpectedly by external magnetic field. Moreover, the result shows that the electron transport through the dot can
be controlled by adjusting the magnetic fields with the aid of the voltage of the leads. The equilibrium-charge diagram is a compact tool
to show clearly the condition of the electron transport, which helps to use single quantum dot as a magnetic electron-transport switch.

\section{Acknowledgments}
This work was supported by the National Natural Science Foundation of China (Grant No 10674084).

\end{document}